\documentclass[aps,prl,letterpaper,twocolumn,preprintnumbers,floatfix,superscriptaddress,letterpaper]{revtex4}
\pdfoutput=1 
\usepackage{graphicx,dcolumn,bm,epsfig,cancel,hyperref}
\usepackage{amsmath}
\usepackage{amssymb, times}

\def\fun#1#2{\lower3.6pt\vbox{\baselineskip0pt\lineskip.9pt
\ialign{$\mathsurround=0pt#1\hfil##\hfil$\crcr#2\crcr\sim\crcr}}}

\def\lsim{\mathrel{\rlap{\raise 2.5pt \hbox{$<$}}\lower 2.5pt\hbox{$\sim$}}}
\def\gsim{\mathrel{\rlap{\raise 2.5pt \hbox{$>$}}\lower 2.5pt\hbox{$\sim$}}}  

\renewcommand{\Im}{{\rm Im\thinspace}}  

\input epsf

\newcommand{\be}{\begin{equation}}
\newcommand{\ee}{\end{equation}}
\newcommand{\bea}{\begin{eqnarray}}
\newcommand{\eea}{\end{eqnarray}}

\usepackage{color}

\begin{document}
\title{Impact of a CP Violating Higgs: from LHC to Baryogenesis}

\author{Jing Shu}
\affiliation{State Key Laboratory of Theoretical Physics and Kavli Institute for Theoretical Physics China (KITPC), Institute of Theoretical Physics, Chinese Academy of Sciences, Beijing 100190, P. R. China}
\author{Yue Zhang}
\affiliation{California Institute of Technology, Pasadena, CA 91125, USA}

\begin{abstract}
We observe a generic connection between LHC Higgs data and electroweak baryogenesis: the particle that contributes to the CP odd $hgg$ or $h \gamma \gamma$ vertex would provide the CP violating source during a first order phase transition. It is illustrated in the 2HDM that a common complex phase controls the lightest Higgs properties at the LHC, electric dipole moments and the CP violating source for electroweak baryogenesis. We perform a general parametrization of Higgs effective couplings and a global fit to the LHC Higgs data. Current LHC measurements prefer a nonzero phase for $\tan \beta \lesssim 1$ and EDM constraints still allow an order one phase for $\tan\beta\sim1$, which gives sufficient room to generate the correct cosmic baryon asymmetry. We also give some prospects in the direct measurements of CP violation in the Higgs sector at the LHC. 
\end{abstract}

\preprint{CAS-KITPC/ITP-365}
\preprint{CALT-68-2926}

\maketitle

\noindent{\bfseries Introduction.} The presence of CP violation (CPV) is always an important aspect in particle physics, which unambiguously leads to discoveries and open questions. In the Standard Model (SM), the CPVs in the K and B-meson systems have established the Cabbibo-Kobayashi-Maskawa (CKM) matrix.
Sakharov~\cite{Sakharov:1967dj} has observed that CPV is essential for creating the apparent asymmetry between matter and anti-matter in our universe. Unfortunately, the CP phase in the CKM matrix is always accompanied with huge suppression from the large quark mass hierarchy when used to generate baryons. Therefore, the search for other sources of CPV would be indispensable for beyond SM physics.

The observation of a SM Higgs-like boson at the Large Hadron Collider (LHC) with a mass at around 125 GeV was announced last summer \cite{lastjuly}. 
Since then, more data has been accumulated~\cite{ATLAS,CMS} and more sophisticated analysis has been carried out based on various Higgs production and decay channels mostly assuming CP conservation~\cite{Giardino:2013bma}, with only few exceptions~\cite{Freitas:2012kw,Celis:2013rcs,Djouadi:2013qya}. 
If CP is violated, both higher dimensional CP even and odd operators would contribute to $gg \rightarrow h$ and $h \rightarrow \gamma \gamma$ processes without interference.
One would expect the results of the Higgs global fits to be different in structure from previous studies. 
Interestingly, the same source of CPV would contribute to fermion electric dipole moment (EDM)~\cite{McKeen:2012av,Chang:2013cia}, and the interplay between the Higgs properties and low energy constrains would be highly non-trivial.  

The CPV source manifests in the higher dimensional Higgs and EDM operators can be mediated by a weak-scale particle (fermion) $X$ with sizable Higgs couplings.
We point out this has an intrinsic connection to electroweak baryogenesis (EWBG) in the early universe. To see this more clearly, consider the renormalizable couplings of $X$ to the Higgs boson, which can be generically parametrized as
$m \bar X [1 + c_X h/v + (\xi + \tilde c_X h/v) i \gamma_5] X$, where $\xi$ is a phase from spontaneous CP violation.
Up to linear terms in $h$ and $\xi$, one can remove the $i \gamma_5$ term by a field redefinition at the expense of generating $\bar\theta$-like terms which linearly depend on $h$ and $\xi$,
\begin{eqnarray}\label{Eq:Connection}
\sim ( \xi + {\tilde  c_X h}/{v} ) F\tilde F \ ,
\end{eqnarray}
where $F$ is the field strength of the gauge symmetry under which $X$ has a charge. In the early universe, during a strongly first order electroweak phase transition (SFO EWPT), $\xi$ can be space-time dependent through the bubble wall. For $SU(2)_L$, the first term in Eq. (\ref{Eq:Connection}) creates a chemical potential and generates a net charge asymmetry $Q_X$:
$\xi(x) F\tilde F \sim \partial_t \xi(x) \cdot Q_{\rm X}$, 
which is nothing but the CPV source for EWBG and $Q_X$ will be furthered reprocessed into baryon asymmetry ($B$) through weak sphaleron transitions. At zero temperature,
the second term in Eq. (\ref{Eq:Connection}) contributes to the CPV $h\to\gamma\gamma$ decay, or $g g \rightarrow h$ production if $X$ is colored. 
The most familiar examples of $X$ include the top and gaugino-Higgsinos. 
It is definitely appealing if baryogenesis can be explained with the knowledge of electroweak scale physics.
After the Higgs discovery
, we enter a territory to measure or constrain the possible CPV sources responsible for $B$ in our universe. 

In this letter, we perform a first study on the direct connection between the latest LHC results on Higgs properties and the baryon number generation from a common CPV phase. We work in a Two-Higgs-Doublet Model (2HDM) and the CPV mediator $X$ is identified as the top quark. 
We study the case when the lightest Higgs boson, with a mass 125 GeV, is a mixture of CP even and odd states. We derive the modified Higgs coupling to other SM particles, and perform a global fit to the current data and extract the constraints on such a phase, which is still allowed to be nonzero, and even favored to be large with $\tan\beta\lesssim1$. We study the electron and neutron EDMs and find the constraints on the same CP phase can be alleviated due to a cancellation with $\tan\beta \sim 1$.
We show such a CP phase is capable of providing all the essential ingredients for EWBG. 
The future advances in precise measurements of Higgs properties, EDMs and refinements in EWBG calculations are anticipated to offer further interplays and pave the way for the genuine origin of CPV for 
baryon asymmetry.

\smallskip
\noindent{\bfseries 2HDM and Sources of CP Violation.} \ \ 
To be specific, we consider the type-II 2HDM, with the Higgs potential
\begin{eqnarray}                    \label{Eq:gko-pot}
\!\!V\!\!&=&\!\!\frac{\lambda_1}{2}(\phi_1^\dagger\phi_1)^2
+\frac{\lambda_2}{2}(\phi_2^\dagger\phi_2)^2
+\lambda_3(\phi_1^\dagger\phi_1) (\phi_2^\dagger\phi_2)  \\
\!\!&+&\!\!\lambda_4(\phi_1^\dagger\phi_2) (\phi_2^\dagger\phi_1) +\frac{1}{2}\left[\lambda_5(\phi_1^\dagger\phi_2)^2+{\rm h.c.}\right] \nonumber \\
\!\!&-&\!\!\frac{1}{2}\left\{m_{11}^2(\phi_1^\dagger\phi_1)
+\!\left[m_{12}^2 (\phi_1^\dagger\phi_2)+{\rm h.c.}\right]\!
+m_{22}^2(\phi_2^\dagger\phi_2)\right\},\!\!\!\!\! \nonumber 
\end{eqnarray}
where $\phi_{1,2}$ are the two Higgs doublets. The tree-level flavor-changing neutral currents can be suppressed by imposing a $Z_2$ symmetry~\cite{Glashow:1976nt} ($\phi_1 \rightarrow -\phi_1$ and $\phi_2 \rightarrow \phi_2$) which is softly broken by $m_{12}$. 
The only complex parameters are $\lambda_{5}$ and $m_{12}^2$ and we can set $\lambda_5$ real by proper rotation of $\phi_{1,2}$ phases.
The corresponding Yukawa couplings respecting the $Z_2$ are
\begin{equation} \label{Eq:Yukawa}
\mathcal{L}_Y = \bar Q_L Y_D \phi_1 D_R + \bar Q_L Y_U (i \tau_2) \phi_2^* U_R + \bar L_L Y_E \phi_1 E_R \ ,
\end{equation}
where $D_R$ or $E_R$ ($U_R$) is defined to be odd (even) under the $Z_2$.
The Higgs vacuum expectation values (VEV) are generally complex, with a relative phase $\xi$,
$\langle \phi_1 \rangle = \left( 0, v\cos\beta/\sqrt{2} \right)^T$,  
$\langle \phi_2 \rangle = \left( 0, v\sin\beta e^{i\xi}/\sqrt{2} \right)^T$.
The minimum condition of the potential solves $\xi$ from the phase of $m_{12}^2$ (recall $\lambda_5$ is real): $\Im (m_{12}^2 e^{i\xi})= (\lambda_5 \sin2\xi) v^2 \sin\beta\cos\beta$
which means there exists one independent physical CP phase.

In this model, the source of CPV arises from the neutral Higgs sector (we define $\sqrt{2}\phi_1^0 = H_1^0 + i A_1^0$, $e^{-i\xi}\sqrt{2}\phi_2^0 = H_2^0 + i A_2^0$ with $H_i^0$, $A_i^0$ being real fields).
Namely, the physical CP-odd state $A^0 = - \sin\beta A_1^0 + \cos\beta A_2^0$
will mix with the even states $H_1^0$, $H_2^0$. 
The mass square matrix ${\cal M}$ in the basis of $(H_1^0, H_2^0, A^0)$ is diagonalized with a real orthogonal $R$, defined as
$R{\cal M}R^{\rm T}={\rm diag}(M_{h_1}^2,M_{h_2}^2,M_{h_3}^2)$
\begin{eqnarray}     \label{Eq:R-angles}
R=\begin{pmatrix}
-s_{\alpha}c_{\alpha_b} & c_{\alpha}c_{\alpha_b} & s_{\alpha_b} \\
s_{\alpha}s_{\alpha_b}s_{\alpha_c} - c_{\alpha}c_{\alpha_c} & -s_{\alpha}c_{\alpha_c} - c_{\alpha}s_{\alpha_b}s_{\alpha_c} & c_{\alpha_b}s_{\alpha_c} \\
s_{\alpha}s_{\alpha_b}c_{\alpha_c} + c_{\alpha}s_{\alpha_c} & s_{\alpha}s_{\alpha_c} - c_{\alpha}s_{\alpha_b}c_{\alpha_c} & c_{\alpha_b}c_{\alpha_c}
\end{pmatrix}
\end{eqnarray}
with $c_\alpha=\cos\alpha$, $s_\alpha=\sin\alpha$. 
In the CP conserving limit, $\alpha_{b,c} \to 0$. In the decoupling limit of second doublet, $\alpha \to \beta - \pi/2$ and $\alpha_{b,c} \to 0$.
The lightest neutral scalar $h_1$, taken to be the SM-like Higgs, with mass $M_1=125$\,GeV, is the following linear combination~\cite{2nd},
$h_1 = -\sin\alpha \cos\alpha_b H_1^0 + \cos\alpha \cos\alpha_b H_2^0 + \sin\alpha_b A^0$.
Using the Yukawa coupling structure in Eq.~(\ref{Eq:Yukawa}), we obtained the couplings of $h_1$ to fermions
\begin{eqnarray} \label{Eq:hff}
\!\!\mathcal{L}_{h_1 f \bar f}&\!\!=\!\!&\frac{m_t}{v} h_1 \bar t \left( c_t +  i \tilde{c}_t\gamma_5 \right) t +  \frac{m_b}{v} h_1 \bar b \left(  c_b +  i \tilde{c}_b \gamma_5  \right) b,
\end{eqnarray}
where $c_t = {\cos\alpha}\cos\alpha_b / {\sin\beta}$, $c_b = -{\sin\alpha} \cos\alpha_b / {\cos\beta}$, $\tilde c_t = - \cot\beta \sin\alpha_b$ and $\tilde c_b = - \tan\beta \sin\alpha_b$. 
The interactions with gauge bosons $WW$ and $ZZ$ are 
\begin{eqnarray} \label{Eq:hvv}
\mathcal{L}_{h_1 VV} = \cos \alpha_b \sin(\beta-\alpha) \mathcal{L}_{h VV}^{\rm SM} \equiv a \mathcal{L}_{h VV}^{\rm SM} \ .
\end{eqnarray}

It is worth pointing out that the CPV coupling of $h_1$ only depends on $\alpha_b$, and is closely connected to the phase $\xi$. 
In order to make their relation more transparent, consider the case $m_{h_2}\approx m_{h_3}\gg m_{h_1}$, we find approximately
$\tan\alpha_b \approx {- \lambda_5 \sin2\xi\,v^2}/[{m_{h^+}^2+(\lambda_4-\lambda_5\cos2\xi) v^2/2}]$,
where $h^+$ is the physical charged Higgs state.
With the second doublet near the weak scale, 
we would expect $\alpha_b \lesssim \xi$.
This is the key relation that motivates our study below. 
The angle $\alpha_b$ are constrained by the Higgs property and the electric dipole moment experiments, 
while the phase $\xi$ is closely connected to the essential CPV source for EWBG.

\begin{figure}[t]
\centerline{\hspace{3mm}\includegraphics[width=1.0\columnwidth]{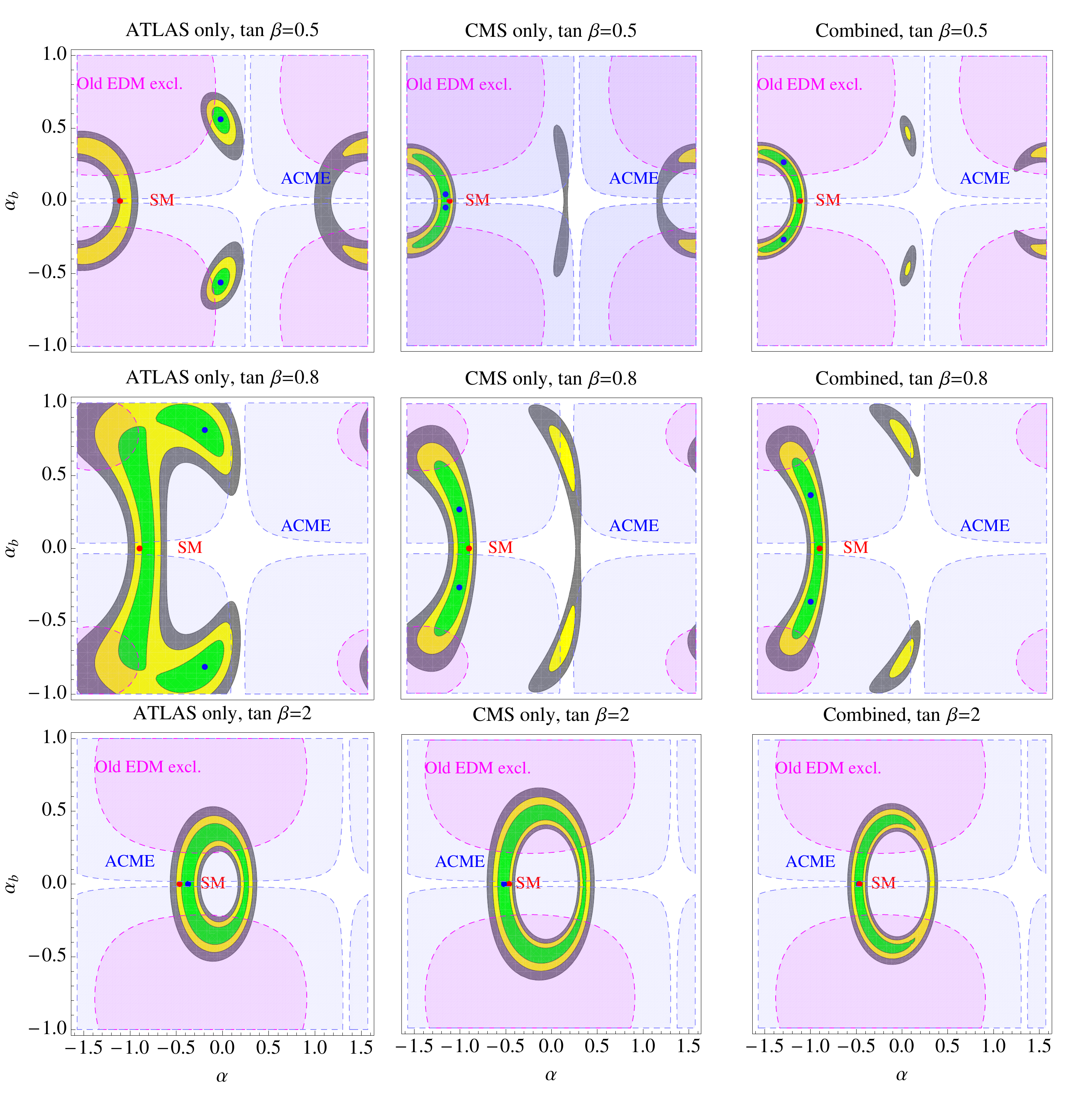}\vspace{-0.3cm}}
\caption{Global fits to the Higgs data for various values of $\tan\beta$. The global minima still prefers a non-vanishing $\alpha_b$ for $\tan\beta\lesssim1$. The magenta region is excluded by electron EDM constraint before Nov. 2013. The latest ACME~\cite{ACME} exclusion ($d_e<10.25\times10^{-29} e$\,cm at 95\% CL) is given by the blue region which is a much stronger constraint.
\vspace{-0.5cm}}\label{Fig:GF}
\end{figure}

\smallskip
\noindent{\bfseries Higgs Properties as Indirect Probe.}  \ \
From the derived interactions (\ref{Eq:hff}) and (\ref{Eq:hvv}), we can obtain the modified Higgs production and decay rates at the LHC. 
The Higgs production via gluon fusion process could happen through both $h_1 GG$ and $h_1G\widetilde G$ operators in an incoherent way, after integrating out the CP conserving and violating $h_1 t\bar t$, $h_1 b\bar b$ interactions. The ratio of the two cross sections is~\cite{Djouadi:2005gj,Carmi:2012in}
\begin{eqnarray} \label{ggh}
\frac{\sigma_{gg\to h_1}}{\sigma^{\rm SM}_{gg\to h_1}} = \frac{(1.03c_t - 0.06 c_b)^2 + (1.57\tilde c_t - 0.06 \tilde c_b)^2}{(1.03 - 0.06)^2},
\end{eqnarray}
for $m_{h_1}=125\,$GeV, and the production cross sections of $h_1$ via $W, Z$ boson fusion and in association with $W, Z$ are simply rescaled from the SM case by
${\sigma_{VV\to h_1}}/{\sigma^{\rm SM}_{VV\to h}} = {\sigma_{V h_1}}/$ ${\sigma^{\rm SM}_{V h}} = a^2$. The heavy Higgs contributions are negligible.

The decay rates into gauge bosons are rescaled by
${\Gamma_{h_1 \to WW}}/{\Gamma_{h \to WW}^{\rm SM}} = {\Gamma_{h_1 \to ZZ}}/{\Gamma_{h \to ZZ}^{\rm SM}} =a^2$.
The decay rates into light fermions are approximately 
${\Gamma_{h_1 \to b\bar b}}/{\Gamma_{h \to b \bar b}^{\rm SM}} = {\Gamma_{h_1 \to \tau^+\tau^-}}/{\Gamma_{h \to \tau^+\tau^-}^{\rm SM}} \approx  c_b^2 + \tilde c_b^2$, by neglecting the final state masses. Similar to the gluon fusion case, the diphoton decay can be separated into CP conserving and violating parts
\begin{eqnarray} \label{Ggaga}
\frac{\Gamma_{h_1 \to \gamma\gamma}}{\Gamma_{h \to \gamma\gamma}^{\rm SM}} = \frac{(0.23 c_t - 1.04 a)^2+(0.35 \tilde c_t)^2}{(0.23 - 1.04)^2} \ .
\end{eqnarray}
Finally, for calculating the Higgs total decay width, the decay to gluons is ${\Gamma_{h_1 \to gg}}/{\Gamma_{h \to gg}^{\rm SM}}={\sigma_{gg\to h_1}}/{\sigma^{\rm SM}_{gg\to h_1}}$.

We make a global fit to the inclusive LHC Higgs data published in March 2013~\cite{ATLAS,CMS}, taking into account the possibility of CPV in the Higgs sector. The most significant change in the latest data is that CMS is no longer seeing an excess in the diphoton channel, while it still persists in the ATLAS result. Therefore, we decided to show both the separate and the combined fits to the ATLAS and CMS data. The best fit points in the effective coefficients $a$, $c_t$, $\tilde{c}_t$, $c_b$, $\tilde{c}_b$ and the 2HDM parameter $\alpha$, $\alpha_b$ for $\tan \beta = 0.8$ are presented in Table~\ref{table:best}. A more comprehensive analysis on the 2HDM parameter $\alpha$, $\alpha_b$ and $\tan \beta$ which includes the exclusion region from the EDM constraints (see below) is shown in Fig.~\ref{Fig:GF}. We find that SM always gives the best fit for $\tan\beta>1$. For $\tan\beta\lesssim1$, better fit points are found with a non-vanishing CP phase $\alpha_b$.


In the presence of CPV Higgs couplings with the top quark $\tilde c_t$, incoherent contributions in Eqs.~(\ref{ggh}) and (\ref{Ggaga})
can modify both the production $gg\to h_1$ and the $h_1\to\gamma\gamma$ decay rate~\cite{Kobakhidze:2012wb}. 
For smaller $\tan\beta\lesssim1$ and $\alpha\approx0$, larger $\tilde c_t$ and smaller $c_b$, $\tilde c_b$ can be achieved simultaneously, with an order one CP phase $\alpha_b$.
The resulting signal strengths are characterized by an enhanced diphoton rate, and a suppressed $Vb\bar b$ rate, both favored by ATLAS (see the first column of Fig.~\ref{Fig:GF}).
The common features of such minima include: 1) enhanced effective $hgg$ coupling, 2) suppressed $\tilde c_t$, $\tilde c_b$, $a$ couplings, and the effective $h\gamma\gamma$ coupling, 3) reduced Higgs total width. These effects are optimized for $\tan\beta\sim1$.
On the other hand, the signals observed by CMS are SM-like. Therefore, the best fit point always lies close to SM. For $\tan\beta<1$, a nonzero $\alpha_b$ gives better fit, because it can accommodate the slight suppression in the $WW$ and $ZZ$ channel as observed by CMS.

\begin{table}[t]
\begin{tabular}{c|cc|ccccc}
\hline
\hline
&   &  & $c_t$ & $\tilde c_t$ & $c_b$ & $\tilde c_b$ & $a$ \\
   & \raisebox{1.2ex}[0pt]{$\alpha$} & \raisebox{1.2ex}[0pt]{$|\alpha_b|$} & $R_{\gamma\gamma}$ & $R_{WW}$ & $R_{ZZ}$ & $R_{Vbb}$ & $R_{\tau\tau}$ \\
\hline
   &              &             & $1.08$ & $-0.91$ & $0.17$ & $-0.58$ & $0.52$ \\
\raisebox{1.2ex}[0pt]{ATLAS}   & \raisebox{1.2ex}[0pt]{$-0.19$} & \raisebox{1.2ex}[0pt]{$0.81$} & 1.35 & 1.28 & 1.28 & 0.47 & 1.71 \\
\hline
 &              &                & $0.83$ & $-0.33$ & $1.04$ & $-0.21$ & $0.96$ \\
\raisebox{1.2ex}[0pt]{CMS}   & \raisebox{1.2ex}[0pt]{$-1.00$} & \raisebox{1.2ex}[0pt]{$0.27$} & 0.91 & 0.83 & 0.83 & 0.93 & 1.02 \\
\hline
 &              &                & $0.82$ & $-0.45$ & $1.00$ & $-0.29$ & $0.93$ \\
\raisebox{1.2ex}[0pt]{Combined}   & \raisebox{1.2ex}[0pt]{$-0.99$} & \raisebox{1.2ex}[0pt]{$0.37$} & 1.05 & 0.86 & 0.86 & 1.02 & 1.18 \\
\hline
\hline
\end{tabular}
\caption{Best fit points with $\tan\beta=0.8$. ATLAS: $\chi^2_{\rm min}-\chi^2_{\rm SM}=-3.27$. CMS: $\chi^2_{\rm min}-\chi^2_{\rm SM}=-1.74$. Combined: $\chi^2_{\rm min}-\chi^2_{\rm SM}=-0.39$. A nonzero CP violating phase is welcomed by the data. \vspace{-0.5cm}}\label{table:best}
\end{table}

\smallskip
\noindent{\bfseries Electric Dipole Moments} \ \
The mixing $\alpha_b$ between the CP even and odd Higgs states leads to a series of low-energy CPV variables, among which we find the EDM of electron gives the leading constraints.
The dominant contribution to electron EDM comes from the Barr-Zee type diagrams at two loop~\cite{Barr:1990vd}.
The lightest Higgs boson can mediate CPV from the top quark and $W$ loops to the electron line~\cite{Gunion:1990iv, Chemtob:1991vv},
\begin{eqnarray}
\left[\frac{d_e}{e}\right]_{i} = \frac{\sqrt{2}\alpha G_F m_e}{(4\pi)^3} F_{i}(z_{i}) \ , 
\end{eqnarray}
where $F_i(z_i)=(16/3) (f( z_i)\tan^2\beta {\rm Im} Z_2 - g (z_i) \cot^2 \beta $ $ {\rm Im} Z_1)$ for top quark and $(6 f ( z_w ) + 10 g( z_w )) (\sin^2\beta \tan^2\beta $ $ {\rm Im} Z_2 + \cos^2\beta  {\rm Im} Z_1 )$ for $W$ boson, $z_i=m_i^2/M_{h_1}^2$ and the loop functions $f(z)$ and $g(z)$ can be found in~\cite{Barr:1990vd}. The CPV variables ${\rm Im} Z_1$, ${\rm Im} Z_2$~\cite{Weinberg:1990me} can be expressed in terms of $c_e$, $\tilde c_e$, $c_t$, $\tilde c_t$, $a$ in Eqs.~(\ref{Eq:hff}) and (\ref{Eq:hvv}) for the Higgs global fits, $\tan^2\beta {\rm Im}\, Z_2 = - \tilde c_b c_t$, $\cot^2\beta {\rm Im} Z_1 =\tilde c_t c_b$ and $(\sin^2\beta \tan^2\beta {\rm Im} Z_2 + \cos^2\beta  {\rm Im} Z_1 ) = a \, \tilde c_b$. 
We neglect the contribution with a Z-boson connecting the electron line and top loop which is always less than 10\%~\cite{Chang:1990sf}. 

\begin{figure}[t]
\centerline{
\includegraphics[width=1.0\columnwidth]{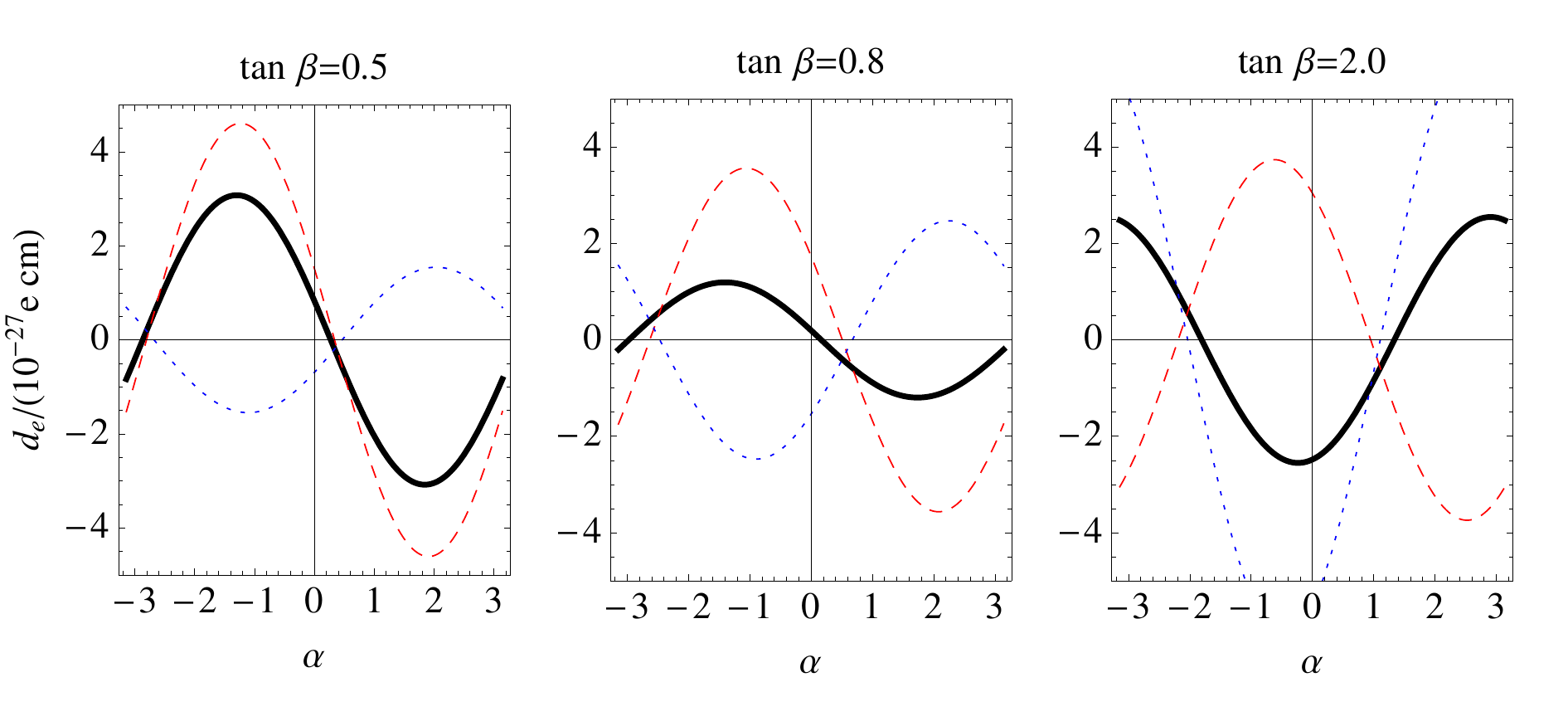}\vspace{-0.4cm}}
\caption{Electron EDM as a function of the angle $\alpha$, for $\tan\beta=0.5, 0.8, 2$ respectively, and with $\alpha_b=0.5$ fixed. The dashed (dotted) curves corresponds to the virtual top-loop ($W$-loop) contribution.\vspace{-0.4cm}}\label{num}
\end{figure}

We find most of the time the top and $W$-loop contributions to electron EDM have opposite signs, and can be minimized simultaneously near $\alpha\approx\beta$. 
The magnitude of the $W$-loop part is more sensitive to $\tan\beta$, and the maximal cancellation happens near $\tan\beta\sim1$. In these regimes, the electron EDM limit can be satisfied without suppressing the CPV phase $\alpha_b$. These features are illustrated in Fig.~\ref{num}. We use the 95\% confidence level limit of the latest electron EDM measurement~\cite{Hudson:2011zz}, $d_e < 1.25\times10^{-27}\,e\,$cm.
The exclusion in $\alpha-\alpha_b$ plan is shown as the magenta region in Fig.~\ref{Fig:GF}. The neutron EDM constraint~\cite{Baker:2006ts} from valence quark EDM, chromoelectric dipole moment and the Weinberg operators are also considered. We find that they do not impose any relevant constraint comparing to the electron EDM. 

So far we have neglected the charged and other neutral Higgs states in both the Higgs fit and the EDM calculation. 
The charged Higgs mass is constrained by $b\to s\gamma$ transition~\cite{Misiak:2006zs} and the measurement of the $R_b$ at LEP~\cite{Haber:1999zh,WahabElKaffas:2007xd} to be heavier than 300--400\,GeV. The bound can be significantly stronger if $\tan \beta$ is far less than one \cite{WahabElKaffas:2007xd, Barroso:2012wz}. 

\smallskip
\noindent{\bfseries Electroweak Baryogenesis.} \ \
During a SFO EWPT, the top quark mass has a space-time dependent phase varying across the bubble wall. This results in a CPV source (for bottom quark case, see~\cite{Liu:2011jh}) which can be estimated as~\cite{Huet:1995sh,Lee:2004we}
\begin{eqnarray}
S_t(z) \approx \frac{3}{2\pi^2} \left( \frac{m_t}{v \sin\beta} \right)^2 v_T^2(z) \theta'(z) v_w T \ ,
\end{eqnarray}
where we take $L_w=5/T$, $v_w=0.02$, and $z<0$ $(>0)$ corresponds to unbroken (broken) side of the expanding bubble.
We assume the following shapes of the bubble wall and the complex phase, $v_T/T = \zeta_T [1+\tanh(z/L_w)]/2$ 
and $\theta(z) = \theta_{\rm brk} - \Delta \theta [1-\tanh(z/L_w)]/2$~\cite{Cline:1995dg}, where
$\Delta\theta$ is the change in the VEV's phase across the wall. To get a SFO EWPT, $\zeta_T\gtrsim\mathcal{O}(1)$,
the heavier neutral scalars are found to be heavier than 400\,GeV~\cite{Fromme:2006cm,Chowdhury:2011ga, Huang:2012wn, Dorsch:2013wja}, 
which is convenient to accommodate in the 2HDM with LHC data \cite{Dorsch:2013wja}.

The imbalance between particle and antiparticle number densities caused by CPV prevails among the quarks and Higgs fields, through the top Yukawa interaction, mass term and strong sphaleron processes. This results in a net asymmetry in the left-handed fermion charge density $n_L$ suppressed by $\Gamma_{ss}$~\cite{Giudice:1993bb}. Here we follow the semi-analytical calculation developed in~\cite{Huet:1995sh, Lee:2004we}.
The relevant thermal rates are 
$\Gamma_{ss} =16 \kappa \alpha_s^4 T$ (with $\kappa=20$ from~\cite{Giudice:1993bb}), $\Gamma_{ws} =120 \alpha_w^5 T$, $\Gamma_h(z) \approx \Gamma_m(z) = (3/2\pi^2) \left({m_t}/{v \sin\beta} \right)^2 v_T^2(z)/T$. In the unbroken phase near the bubble, the weak sphaleron process converts $n_L$ into baryon asymmetry, which is estimated as
\begin{eqnarray}
n_b = - \frac{3 \Gamma_{ws}}{2v_w} \int_{-\infty}^0 n_L(z) e^{15 \Gamma_{ws} z/(4 v_w)} dz \ .
\end{eqnarray}

\begin{figure}[t]
\centerline{
\includegraphics[width=0.9\columnwidth]{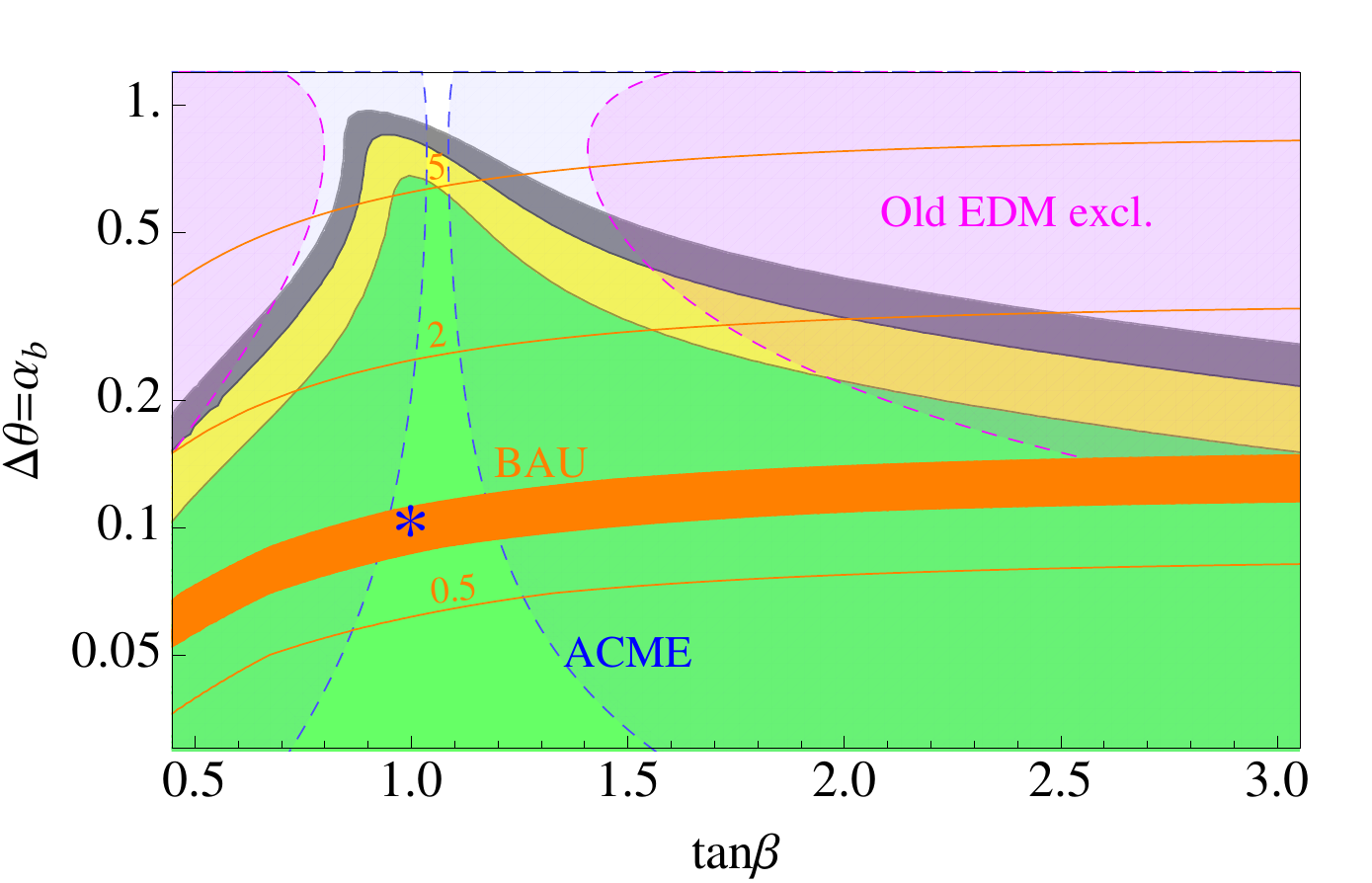}\vspace{-0.3cm}}
\caption{Values of $\Delta\theta$ and $\tan\beta$ consistent with the observed $\eta_b$. We take the constraints from Fig.~\ref{Fig:GF} with $\alpha=\beta-\pi/2$, and $\Delta\theta=\alpha_b$. The blue star point is the benchmark point from the numerical studies in~\cite{Fromme:2006cm} that $\Delta \theta = \xi \approx 0.1$ for $m_{h_1}= 125$ GeV and $m_{h_2}$ = 400 GeV. 
\vspace{-0.6cm}}
\label{bau}
\end{figure}

We find the observed baryon asymmetry to entropy density ratio $\eta_b = {n_b}/{s}\approx (0.7-0.9)\times 10^{-10}$~\cite{pdg,Ade:2013lta} can be obtained with $\Delta\theta$ around 0.2 ($\zeta_T =1.5$). The value of $\Delta \theta$ is solved numerically from the Higgs potential in Eq. (\ref{Eq:gko-pot}). It has been shown that $\Delta\theta$ is of similar size to the zero temperature phase $\xi$ for small $m_{h_1}$~\cite{Fromme:2006cm,Cline:2011mm} close to $m_{h_1} = 125$ GeV. Therefore it is convincing to assume $\Delta \theta \approx \alpha_b$ and present the corresponding connection to EDM and Higgs global fits in Fig.~\ref{bau}. 
We can see that successful EWBG sets a lower bound on the CPV phase. Such a phase will keep being probed directly or indirectly in the future LHC Higgs measurements and low energy experiments like EDM, and being used to test the viability of the EWBG scenario. We do notice though the final baryon number density is sensitive to the choices of $v_w$, $\zeta_{T}$ and $\Gamma_{ss}$, etc.. 
A more precise calculation of $\eta_b$ would require improved determination of these quantities which involves higher-order and non-perturbative calculations.

\smallskip
\noindent{\bfseries Direct Probe of CP Violation in Higgs sector.} \ \ Here we briefly discuss the prospects of measuring the CPV in the Higgs sector. The $h\to ZZ^*\to4\ell$ process has been used to constrain the CP odd coupling to $Z$-boson~\cite{CMS2, Cao:2009ah, Boughezal:2012tz}). Nevertheless, a relevant limit can not be obtained to any model since the observable is suppressed by the large tree level CP even coupling. The physical effects from CP odd and even operators in the $2 \gamma$ or $Z \gamma$ channel are comparable but such discriminations require better identification of photon polarization in the Bethe-Heitler process~\cite{HBH}. A more promising channel could be the gluon fusion production of $h$ in together with two forward jets~\cite{Klamke:2007cu} or the $t \bar{t} h$ production~\cite{Gunion:1996xu}.
The virtual effect of a CPV Higgs coupling can also be probed in the top pair production with the leptonic decay channel~\cite{Schmidt:1992et}.
We leave a systematic classification and quantitative study of these signatures employing the LHC data for a future work.

\smallskip
\noindent{\bfseries Conclusion.} \ \ 
In summary, we elaborate the connection between the LHC data and EWBG in a 2HDM. 
We performed a global fit to the latest LHC Higgs data, and find a nonzero CP phase is favored for $\tan\beta\lesssim1$. When combined with the electron and neutron EDM constraints, we find the phase is allowed to be as large as order one near $\tan\beta\sim1$. We show that this phase can provide the CPV source for EWBG. The future improvements in measuring the Higgs properties at LHC and the EDMs would enable us to probe the possible origin of CPV beyond the CKM matrix, that is connected to the baryon asymmetry of the universe.

\smallskip
\noindent{\bfseries Acknowledgements.} \ \  We thank Q.H.~Cao, C.~Cheung, T.\, Liu, M.~Ramsey-Musolf, M.B.~Wise for useful discussions and especially J.~Zupan for the collaboration at the early stage. YZ acknowledge the hospitality from Institute of Theoretical Physics and Kavli Institute for Theoretical Physics China where part of the work has been done. The research of YZ is funded by the Gordon and Betty Moore Foundation through Grant No.\,776 to the Caltech Moore Center for Theoretical Cosmology and Physics, and in part by the DOE Grant DE-FG02-92ER40701.



\begin{thebibliography}{40}
\bibitem{2nd}
{The second lightest state could also serve as the 125 GeV resonance, but it gets severely constrained from flavor and precision electroweak physics, therefore we do not consider this possibility throughout the paper.}
\end{thebibliography}
\end{document}